\documentclass[prl,twocolumn,a4paper,groupedaddress,showpacs,floatfix,aps,10pt,longbibliography,nobibnotes]{revtex4-2}

\usepackage{graphicx,amsmath,amssymb,amsfonts,dsfont,subfigure,color}

\begin{document}

\title{Telecom-wavelength spectra of a Rydberg state in a hot vapor}

\author{Wenfang Li${}^{1}$}
\thanks{The first two authors contributed equally.}
\email{wendili@nus.edu.sg}
\author{Jinjin Du${}^{1}$}
\thanks{The first two authors contributed equally.}
\author{Mark Lam${}^{1}$}
\author{Wenhui Li${}^{1}$}
\email{wenhui.li@nus.edu.sg}

\affiliation{Centre for Quantum Technologies, National University of Singapore, 3 Science Drive 2, Singapore 117543${}^1$}




\begin{abstract}
We study telecom-wavelength spectra of a Rydberg state in an atomic vapor with a three-photon excitation scheme. Two lasers of 780 nm and 776 nm are used to pump Rubidium-85 atoms in a vapor cell to the $5D_{\mathrm{5/2}}$ state, from which a probe beam of 1292 nm in the O-band telecommunication wavelength drives a transition to the $21F_{\mathrm{7/2}}$ Rydberg state. We investigate the probe spectra over the power of pump lasers. The simulation based on a 4-level theoretical model captures the main features of the experimental results. This spectroscopic study paves the way for future experiments of making a direct link between fiber optics and radio transmission via Rydberg atoms.
\end{abstract}


\maketitle


Atomic media, with their near-resonant coherent coupling to laser fields, give rise to some very unique optical responses, such as electromagnetically induced transparency (EIT) as well as large high-order non-linearity~\cite{fleischhauer:05}. The seminal work on optical detection of Rydberg states via EIT~\cite{mohapatra2007coherent} has opened up numerous opportunities to utilize the exaggerated properties of Rydberg atoms for nonlinear optical applications of various kinds. On the one hand, the strong dipole-dipole interaction of Rydberg atoms greatly enhances the non-linearity down to the single-photon level, enabling single-photon generation and photon-photon quantum gate~\cite{dudin2012strongly,peyronel2012quantum,busche2017contactless,tiarks2019photongate}. On the other hand, high sensitivity of Rydberg atoms to electric field from DC to terahertz (THz) frequencies, leads to strong electro-opto effects and extends nonlinear frequency mixing from optical to microwave and THz domains~\cite{mohapatra2008giant,sedlacek2012microwave,Wade2017}. The resulting applications include electric field sensing for establishing traceable SI standards and radio-over-fiber in communication network~\cite{holloway2017rfmetrology,deb2018radio,meyer2018digital,jing2020}, THz imaging and generation~\cite{downes2020,lam2021directional}, and coherent microwave-to-optical conversion~\cite{han2018coherent,tu2022}. The optical connection between the ground state and a Rydberg state is most commonly done via transitions using two photons, for instance, a 780-nm (852-nm) probe field and a 480-nm (514-nm) coupling field in the case of rubidium (cesium) atoms. Meanwhile, three-photon excitation schemes have been investigated as well, as these schemes have advantages in eliminating Doppler broadening with commonly available diode lasers~\cite{carr2012three,thaicharoen2019,ripka2021}. Notably, all experiments so far have the field driving the transition from the ground state to the first excited states (and, in a handful of cases, to the second excited states), e.g. a laser of 780-nm (852-nm) for rubidium (cesium) atoms, as the optical probe carrying a signal to be transmitted and detected.

\begin{figure}[htb]
\centerline{
\includegraphics[width=0.8\linewidth]{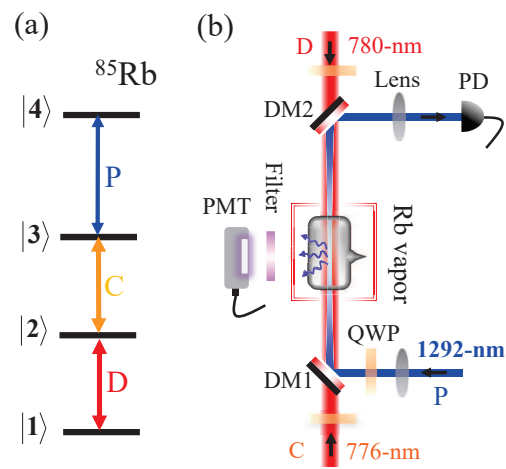}}
 \caption{(a) Atomic energy levels and laser fields driving the respective transitions. (b) Schematic of the experimental cell. QWP: quarter-wave plates, DM1(2): dichroic mirrors, PD: photoreceiver detector, PMT: photo-multiplier tube.}
\label{Fig1}
\end{figure}

In this letter, we investigate telecom-wavelength spectra of a transition from an intermediate state to a Rydberg state with the presence of two pump lasers. Shown in Fig.~\ref{Fig1}(a) is the energy-level scheme. $^{85}$Rb atoms are pumped from the ground state $\left| 1\right\rangle = \left|5S_{1/2}, F=3\right\rangle$ via the first excited state $\left| 2\right\rangle = \left|5P_{3/2}, F=4\right\rangle$ to the intermediate state $\left| 3\right\rangle = \left|5D_{5/2}, F=5\right\rangle$ by two lasers D and C, which have the wavelengths of 780 nm and 776 nm, respectively. A probe laser P of 1292 nm drives the transition to a Rydberg state $\left| 4\right\rangle = \left|21F_{7/2}, J=7/2\right\rangle$. We characterize the spectroscopic features of the probe and investigate their dependence on the power of pumps as well as on the resulting signature of the optically pumped atomic vapor. This is, to our knowledge, the first experiment to study the optical response of a telecom wavelength in a Rydberg medium. Such a response is readily modified when the medium is exposed to a microwave (MW) field that couples the Rydberg state $\left| 4\right\rangle$ to another nearby Rydberg level. Therefore, this spectroscopic study constitutes the first step in establishing a direct link between RF or MW field and optical telecom wavelength via Rydberg media, which can be utilized for applications in communication networks, either classical or quantum.

A sketch of the experimental setup is shown in Fig.~\ref{Fig1}(b). A rubidium vapour of natural abundance is hosted in a cell of length L = 100 mm and of 25.4-mm diameter, and the cell is heated in an oven, the temperature of which is typically set to be {T $\sim$ 80 °C}. The atomic density of the $^{85}$Rb ground state $\left|1\right\rangle$ is measured to be around $n_0 = 2 \times{10}^{11}{\rm cm}^{-3}$. The three laser beams, all of which are derived from external cavity diode lasers, propagate co-axially through the vapor, and their respective $1/e^2$ radii of $w_\text{D}$ = 714 µm, $w_\text{C}$ = 635 µm, and $w_\text{P}$ = 211 µm overlap around the center of the cell. While the two pump beams counter-propagate, the probe beam P propagates along the same direction as that of the pump beam C. The detuning of the laser frequency $\omega_L$ from its respective atomic resonance $\omega_{ij}$, as illustrated in Fig.~\ref{Fig1}(a), is defined as $\Delta_{\text{L}}=\omega_{\text{L}}-\omega_{ij}$, where $\text{L}\in {\{\text{D}, \text{C}, \text{P}\}}, i\in{\{1, 2, 3\}}$, and $j =i+1$.

Spectroscopy of the probe is taken with the frequencies of the two pumps being fixed as follows. The 780-nm pump laser is locked at $\Delta_\text{D}/2\pi = -$60 MHz by a Doppler-free dichroic atomic vapor laser lock. The 776-nm pump laser is stabilized at the two-photon resonance $\Delta_\text{C} = -\Delta_\text{D}$ using a vapor cell (not the one in Fig.~\ref{Fig1}(b)), where a small fraction (a few tens of µW) of each pump beam is sent to counter-propagate in the same way as that shown in Fig.~\ref{Fig1}(b). At the two-photon resonance, atoms are pumped to the $\left|3\right\rangle$ state, and produce a fluorescence emission at a wavelength of 420 nm through a cascade decay~\cite{Akulshin:09,Vernier:10}. This fluorescence is collected by a photo-multiplier tube (PMT, Hamamatsu H10722-01) to generate an error signal for locking the 776-nm laser frequency. Note here that the fluorescence signal is mostly from a fraction of atoms in the hot vapor, which move towards the 780-nm beam with velocity $v_0$, such that $\Delta_\text{C} - k_\text{C} v_0 = \Delta_\text{D} - k_\text{D} v_0 = 0$, where $k_\text{D} $ and $k_\text{C}$ are the wave vectors of the two pumps D and C, respectively. That is, this group of atoms in their rest frame see the two lasers resonantly driving the  $\left|1\right\rangle\rightarrow\left|2\right\rangle$ and  the $\left|2\right\rangle\rightarrow\left|3\right\rangle$ transitions. Since $k_\text{D} \sim - k_\text{C}$, it is appropriate to make the approximation $\Delta_\text{C} = -\Delta_\text{D}$ and $v_0 = + 46$ m/s.

\begin{figure}[tb]
\centerline{
\includegraphics[width=0.85\linewidth]{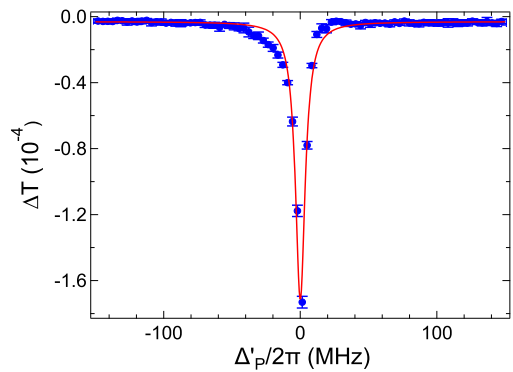}}
\caption{Spectrum of $\Delta T$ versus $\Delta_P'$. The error bars are standard deviation over 10 measurements. The red solid curve shows a fit of a Lorentzian function to the data. The input powers of the pump lasers for the spectrum are respectively $P_\text{D}$ = 50 $\mu$W and $P_\text{C}$ = 100 $\mu$W. The Rabi frequency of the probe P is $\Omega_\text{P}$= 2$\pi\times1.6$ MHz ($\sim$ 100 $\mu$W), which is used throughout the spectroscopic measurements in the paper.  }
\label{Fig2}
\end{figure}

In the experimental cell shown in Fig.~\ref{Fig1}(b), the two pumps, with input power of $P_\text{D}$ and $P_\text{C}$, will most efficiently drive atoms of velocity $v_0$ into the $\left|3\right\rangle$ state, as argued above. Therefore, the detuning of the probe seen by these atoms is effectively $\Delta_\text{P}' = \Delta_\text{P} - k_\text{P}v_0$, where ${k_\text{P}}$ is the wave vector of the probe light. As the effective detuning of the probe $\Delta_\text{P}'$ is scanned through the resonance of the $\left|3\right\rangle\rightarrow\left|4\right\rangle$ transition, the transmitted power of the probe is monitored by a photodiode (Newfocus, 2053-FS-M).  Lock-in method, meanwhile, is applied for acquiring the spectra by modulating the amplitude of the pump C. To record the change in the transmission of the probe from driving the $\left|3\right\rangle\rightarrow\left|4\right\rangle$ transition only, we cancel out the absorption of the probe due to other elements in its path. This is done by measuring the transmitted probe power $P_{\text{P-on}}$ and $P_{\text{P-off}}$ with the pumps on and off respectively and then taking a ratio between them. That is, the change in the probe transmission is defined as $\Delta T = P_{\text{P-on}}/P_{\text{P-off}} - 1$. Given in Fig.~\ref{Fig2} is an example spectrum of the probe's transmission change $\Delta T$ versus the effective detuning $\Delta_\text{P}'$. Under the conditions specified in the caption, the peak change of the probe transmission at resonance $\Delta_\text{P}' = 0$ can reach $\Delta T_0=-1.7\times 10^{-4}$. Fitting of a Lorentzian function to the spectrum gives a full width at half maximum (FWHM) of 8.7 $\pm 0.3$\,$ $\rm{MHz}. This spectral width, much less than the Doppler broadening, shows that the probe sees a sub-Doppler atomic medium, produced by the two counter-propagating pump beams.

\begin{figure}[!b]
\centerline{
\includegraphics[width=\linewidth]{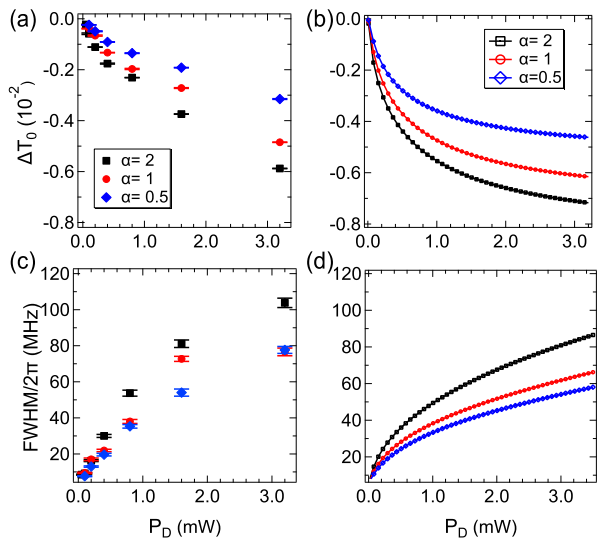}}
 \caption{Resonance transmission change $\Delta T_0$ (a) and FWHM (c) of the probe spectra measured in experiment versus $P_{D}$ at three values of $\alpha$. (b) and (d) are the simulation results corresponding to (a) and (c) respectively. The error bars of experimental data are from fitting spectra.}
\label{Fig3}
\end{figure}

We further investigate how the spectra of the probe depend on the input powers of the two pumps, ${P_\text{D}}$ and ${P_\text{C}}$, as well as their ratio, $\alpha$ = ${P_\text{C}}$/ ${P_\text{D}}$. Spectra are taken under different powers of the pumps as well as three different values of ratio $\alpha$. Their peak transmission change $\Delta T_0$ and FWHM are extracted in the same way as that described in the discussion of Fig.~\ref{Fig2}, and are plotted versus ${P_\text{D}}$ in Figs.~\ref{Fig3} (a) and (c), respectively. It can be clearly seen that higher pump powers give a larger $\Delta T_0$ and a wider FWHM, so does a larger $\alpha$. This trend is generally intuitive to understand, as higher pump powers likely result in a larger population in the $\left|3\right\rangle$ state as well as a power broadening of the state.

To obtain the quantitative understanding about spectroscopic features of the probe, we numerically simulate the probe's transmission through the optically pumped hot atomic ensemble, by theoretically modeling the four-level scheme of Fig.~\ref{Fig1}(a), where all other energy levels are neglected. The probe's transmission change $\Delta T$, as defined above, is related to the susceptibility of the probe $\chi_\text{P}$ as
\begin{equation} \label{deltaT}
\Delta T = {\rm exp}\left(-{\rm Im} \left\{\chi_\text{P}\right\} k_\text{P} {\rm L }\right) - 1.
\end{equation}
The probe's susceptibility, proportional to the coherence of the $\left|3\right\rangle\rightarrow\left|4\right\rangle$ transition $\chi_\text{P} \propto \rho_{34}$, can be calculated by solving the steady-state solution to the master equation describing the quantum dynamics of the single-atom density operator $\rho$, which can be written as
\begin{equation}\label{masterequ}
\partial _{t} \rho=-\frac{i}{\hbar}\left [H,\rho \right]+L\left (\rho\right ),
\end{equation}
where $H$ is the Hamiltonian of a single atom, and $L\left(\rho\right)$ accounts for spontaneous emission of the excited states that are described by standard Lindblad decay terms.

In the electric-dipole and rotating-wave approximation, the single-atom Hamiltonian describing the four atomic levels interacting with the three laser fields is expressed as
\begin{equation}\label{hamiltonian}
\begin{aligned}
H=& -\hbar\Bigg[\delta_\text{D}{\sigma}_{22}+\left(\delta_\text{D}+\delta_\text{C}\right){\sigma}_{33} +\left(\delta_\text{D}+\delta_\text{C}+\delta_\text{P}\right) {\sigma}_{44}\Bigg]\\
& -\frac{\hbar}{2}\left(\Omega_\text{D}{\sigma}_{21}+\Omega_\text{C}{\sigma}_{32}+\Omega_\text{P}{\sigma}_{43}+\mathrm{H.c.}\right),
\end{aligned}
\end{equation}
where ${\sigma}_{ij}=\left|i\right\rangle\left\langle j\right|$ are the atomic transition operators, and $\Omega_\text{L}$ is the Rabi frequency of the laser field L driving the corresponding transition. Here the effective detuning $\delta_\text{L} = \Delta_\text{L}- k_Lv$,  seen by an atom of velocity $v$ along the direction parallel to $k_L$, takes into account the Doppler effect, where $v$ follows the normalized one-dimensional Maxwell distribution $G(v)$. Meanwhile, $L\left(\rho\right)$ is written with the standard Lindblad decay terms as
\begin{equation}\label{lindblad}
\begin{aligned}
L\left(\rho\right)= & -\frac{\Gamma_{2}}{2}\left({\sigma}_{22}\rho+\rho{\sigma}_{22}-2{\sigma}_{12}\rho{\sigma}_{12}^\dag\right)\\
&-\frac{\Gamma_{3}}{2}\left({\sigma}_{33}\rho+\rho{\sigma}_{33}-2{\sigma}_{23}\rho{\sigma}_{23}^\dag\right) \\
&-\frac{\Gamma_{4}}{2}\left({\sigma}_{44}\rho+\rho{\sigma}_{44}-2{\sigma}_{34}\rho{\sigma}_{34}^\dag \right),\\
\end{aligned}
\end{equation}
where $\Gamma_{j}$ is the decay rate of the excited state$\left.\ \left|j\right.\right\rangle$ ( $j \in \left\{2, 3, 4  \right\}$ ).

At a particular $\Delta_\text{P}$ on a spectrum of the probe, atoms with each velocity $v$ see a particular set of $\delta_\text{L}$ in Eq.~(\ref{hamiltonian}), the coherence $\rho_{34}$ can be solved accordingly. Then the overall susceptibility of the probe at this detuning is calculated by integrating over $G(v)$, that is, ${\rm Im} \left\{\chi_\text{P}\right\}= \frac{n_0d_{34}^2 \int{G(v)\rm Im}\left\{\rho_{34}\right\}dv}{\Omega_p\epsilon_0\hbar}$, where $\epsilon_{0}$ is
the vacuum permittivity and $d_{34}$ is the dipole matrix element of the $\left |3\right\rangle\rightarrow\left|4\right\rangle$  transition. Consequently, the probe transmission change $\Delta T$ can be calculated according to Eq.~(\ref{deltaT}). A full spectrum can be obtained as $\Delta_\text{P}$ is scanned over the resonance. Since both $\Delta_\text{D}$ and $\Delta_\text{C}$ are fixed as discussed before, the probe susceptibility mostly comes from atoms with a velocity around $v_0$. Therefore, the spectral response is in the vicinity centered around $\Delta_\text{P}' = 0$, as seen in Fig.~\ref{Fig2}.

Shown in Figs.~\ref{Fig3} (b) and (d) are the simulated resonance transmission change $\Delta T_0$ and FWHM of the probe spectra under the same conditions as that in Figs.~\ref{Fig3} (a) and (c), respectively. The results from experiment and simulation generally agree. The residual discrepancy is likely due to the fact that our simplified four-level model takes into account neither coupling to other nearby energy levels nor various kinds of experimental inhomogeneity. Nevertheless, the simulation captures the main features of the probe spectra. Higher pump powers result in larger transmission change $\Delta T_0$ as well as wider FWHM. This can be qualitatively explained as follows. At higher pump powers, the single-atom density matrix $\rho_{33}$ is larger and the energy level of the $\left|3\right\rangle$ state is power broadened. While the former simply increases the atomic population in the $\left|3\right\rangle$ state within a same velocity group; the latter helps to recruit more atoms over a broader range of velocity around $v_0$ into the $\left|3\right\rangle$ state. Consequently, the spectra at such conditions have larger $\Delta T_0$ as well as wider FWHM. This argument is further supported by the observation that under the same input power $P_\text{D}$, the power ratio $\alpha = 2$ results in larger $\Delta T_0$ as well as wider FWHM than the power ratios $\alpha = 0.5$ and 1. The power ratio $\alpha = 2$ means a higher $P_\text{C}$ that directly drives the $\left|2\right\rangle$ and $\left|3\right\rangle$ transition. Hence more population of the $\left|3\right\rangle$ state are available due to the two aspects discussed above.

\begin{figure}[!t]
\centerline{
\includegraphics[width=0.9\linewidth]{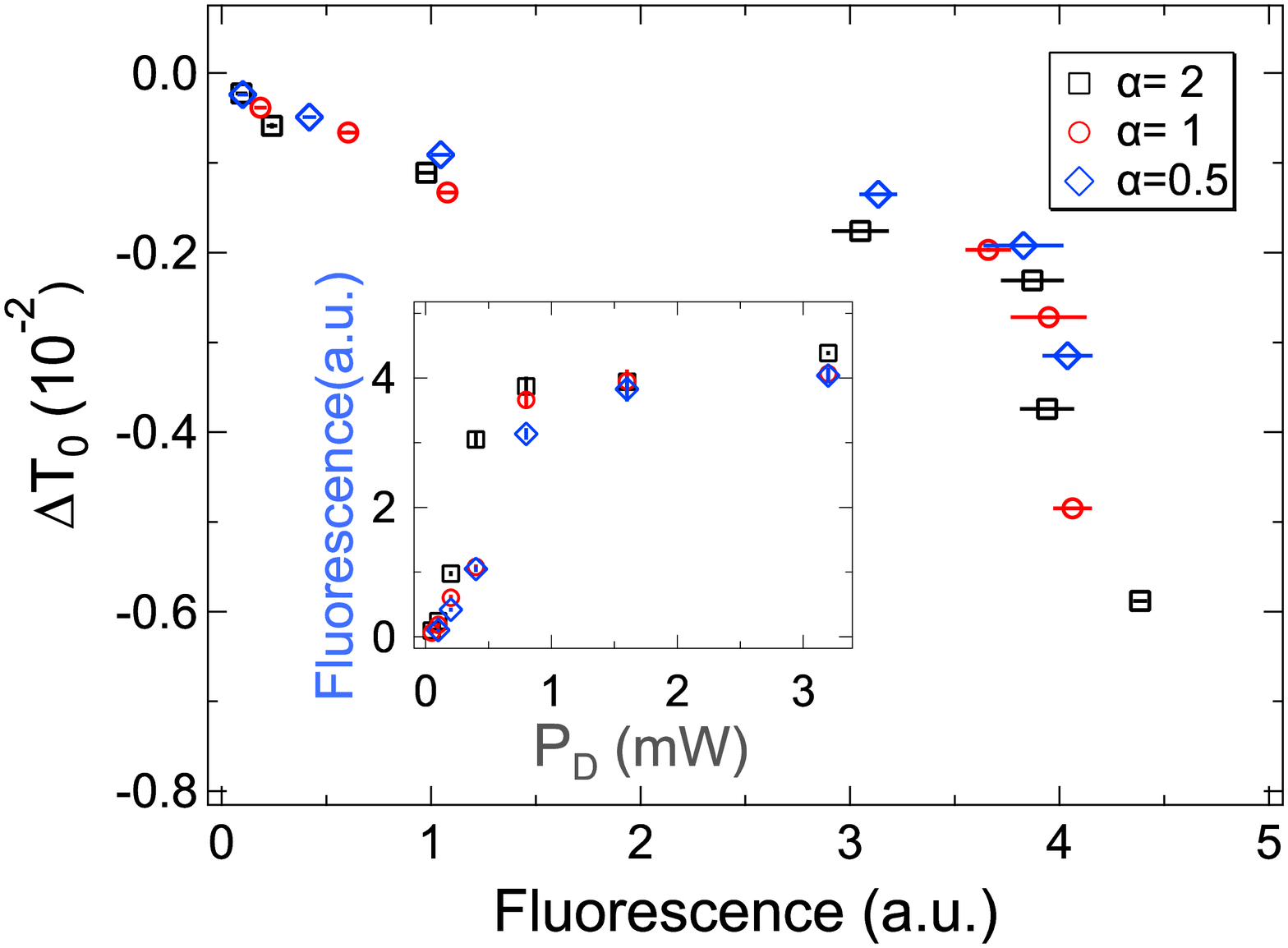}}
 \caption{Resonance transmission change $\Delta T_0$ versus 420-nm fluorescence collected at $P_\text{D}$ and $\alpha$ of Fig.~\ref{Fig3}. The error bars are the same as that in Fig.~\ref{Fig3} (a). The inset figure shows the level of fluorescence versus $P_\text{D}$ for three different values of $\alpha$. The error bars of the inset figure are standard deviation over 20 measurements.}
\label{Fig4}
\end{figure}

Due to the decay from the two excited states $\left|2\right\rangle$ and $\left|3\right\rangle$, spectroscopy of the probe doesn't directly depend on the ground state $\left|1\right\rangle$ or the first excited state $\left|2\right\rangle$ in a coherent way. Instead, it may be appropriate to consider that the probe sees a medium made of the steady-state atomic population in the $\left|3\right\rangle$ state with the presence of the two pumps. To verify this experimentally, we measure the blue fluorescence at 420 nm using PMT, as shown in Fig.~\ref{Fig1} (b), under the pump powers and power ratios used in Fig.~\ref{Fig3}. This fluorescence has a direct proportionality to the population in the $\left|3\right\rangle$ state~\cite{Akulshin:09,Vernier:10}. In Fig.~\ref{Fig4}, the resonance transmission change $\Delta T_0$ in Fig.~\ref{Fig3} (a) is plotted against the measured fluorescence. It can be seen that the different plots from different $\alpha$ in Fig.~\ref{Fig3} (a) all fall into a single curve that depends on the level of fluorescence, equivalently on the population in the state $\left|3\right\rangle$. There is a sudden steep change in $\Delta T_0$ around the fluorescence level of 4 (a.u.), which is mostly due to the saturation of fluorescence at this level. This saturation can be clearly seen in the inset of Fig.~\ref{Fig4}, where the fluorescence is plotted against $P_\text{D}$ for different $\alpha$. As the generation of this fluorescence involves energy levels outside the four-level scheme, a quantitative understanding requires an advanced model including more states than that being considered in the simple simulation presented here, and is subject to further investigation.

In summary, based on a three-photon excitation involving a Rydberg state in a hot rubidium vapor cell, we investigate an atomic optical spectroscopy in a range of an O-band telecom-wavelength. Here the probe is a telecom light that directly couples to a Rydberg state. This is distinctly different from all other experiments employing a similar ladder structure of four energy levels, where the probe of near-infrared or visible wavelength only couples to Rydberg states indirectly via a coupling light. This spectroscopic study is a precursor to experiments of modifying the telecom probe by microwave field via Rydberg states, which will enable direct link of fiber optics and radio transmission via Rydberg atomic vapors. Moreover, telecom-wavelength spectroscopy of Rydberg states may also be useful for the effort of developing microwave-to-optical conversion via multi-wave mixing in Rydberg media towards applications for quantum interface and networks.

\section*{Acknowledgments}
The authors thank Thibault Vogt for helpful inputs, and acknowledge the support by the National Research Foundation, Prime Minister's Office, Singapore and the Ministry of Education, Singapore under the Research Centres of Excellence programme.


%


\begin{thebibliography}{22}%
\makeatletter
\providecommand \@ifxundefined [1]{%
 \@ifx{#1\undefined}
}%
\providecommand \@ifnum [1]{%
 \ifnum #1\expandafter \@firstoftwo
 \else \expandafter \@secondoftwo
 \fi
}%
\providecommand \@ifx [1]{%
 \ifx #1\expandafter \@firstoftwo
 \else \expandafter \@secondoftwo
 \fi
}%
\providecommand \natexlab [1]{#1}%
\providecommand \enquote  [1]{``#1''}%
\providecommand \bibnamefont  [1]{#1}%
\providecommand \bibfnamefont [1]{#1}%
\providecommand \citenamefont [1]{#1}%
\providecommand \href@noop [0]{\@secondoftwo}%
\providecommand \href [0]{\begingroup \@sanitize@url \@href}%
\providecommand \@href[1]{\@@startlink{#1}\@@href}%
\providecommand \@@href[1]{\endgroup#1\@@endlink}%
\providecommand \@sanitize@url [0]{\catcode `\\12\catcode `\$12\catcode
  `\&12\catcode `\#12\catcode `\^12\catcode `\_12\catcode `\%12\relax}%
\providecommand \@@startlink[1]{}%
\providecommand \@@endlink[0]{}%
\providecommand \url  [0]{\begingroup\@sanitize@url \@url }%
\providecommand \@url [1]{\endgroup\@href {#1}{\urlprefix }}%
\providecommand \urlprefix  [0]{URL }%
\providecommand \Eprint [0]{\href }%
\providecommand \doibase [0]{https://doi.org/}%
\providecommand \selectlanguage [0]{\@gobble}%
\providecommand \bibinfo  [0]{\@secondoftwo}%
\providecommand \bibfield  [0]{\@secondoftwo}%
\providecommand \translation [1]{[#1]}%
\providecommand \BibitemOpen [0]{}%
\providecommand \bibitemStop [0]{}%
\providecommand \bibitemNoStop [0]{.\EOS\space}%
\providecommand \EOS [0]{\spacefactor3000\relax}%
\providecommand \BibitemShut  [1]{\csname bibitem#1\endcsname}%
\let\auto@bib@innerbib\@empty
\bibitem [{\citenamefont {Fleischhauer}\ \emph {et~al.}(2005)\citenamefont
  {Fleischhauer}, \citenamefont {Imamo\v{g}lu},\ and\ \citenamefont
  {Marangos}}]{fleischhauer:05}%
  \BibitemOpen
  \bibfield  {author} {\bibinfo {author} {\bibfnamefont {M.}~\bibnamefont
  {Fleischhauer}}, \bibinfo {author} {\bibfnamefont {A.}~\bibnamefont
  {Imamo\v{g}lu}},\ and\ \bibinfo {author} {\bibfnamefont {J.~P.}\ \bibnamefont
  {Marangos}},\ }\bibfield  {title} {\bibinfo {title} {Electromagnetically
  induced transparency: Optics in coherent media},\ }\href@noop {} {\bibfield
  {journal} {\bibinfo  {journal} {Rev. Mod. Phys.}\ }\textbf {\bibinfo {volume}
  {77}},\ \bibinfo {pages} {633} (\bibinfo {year} {2005})}\BibitemShut
  {NoStop}%
\bibitem [{\citenamefont {Mohapatra}\ \emph {et~al.}(2007)\citenamefont
  {Mohapatra}, \citenamefont {Jackson},\ and\ \citenamefont
  {Adams}}]{mohapatra2007coherent}%
  \BibitemOpen
  \bibfield  {author} {\bibinfo {author} {\bibfnamefont {A.~K.}\ \bibnamefont
  {Mohapatra}}, \bibinfo {author} {\bibfnamefont {T.~R.}\ \bibnamefont
  {Jackson}},\ and\ \bibinfo {author} {\bibfnamefont {C.~S.}\ \bibnamefont
  {Adams}},\ }\bibfield  {title} {\bibinfo {title} {Coherent optical detection
  of highly excited rydberg states using electromagnetically induced
  transparency},\ }\href@noop {} {\bibfield  {journal} {\bibinfo  {journal}
  {Phys. Rev. Lett.}\ }\textbf {\bibinfo {volume} {98}},\ \bibinfo {pages}
  {113003} (\bibinfo {year} {2007})}\BibitemShut {NoStop}%
\bibitem [{\citenamefont {Dudin}\ and\ \citenamefont
  {Kuzmich}(2012)}]{dudin2012strongly}%
  \BibitemOpen
  \bibfield  {author} {\bibinfo {author} {\bibfnamefont {Y.~O.}\ \bibnamefont
  {Dudin}}\ and\ \bibinfo {author} {\bibfnamefont {A.}~\bibnamefont
  {Kuzmich}},\ }\bibfield  {title} {\bibinfo {title} {Strongly interacting
  rydberg excitations of a cold atomic gas},\ }\href@noop {} {\bibfield
  {journal} {\bibinfo  {journal} {Science}\ }\textbf {\bibinfo {volume}
  {336}},\ \bibinfo {pages} {887} (\bibinfo {year} {2012})}\BibitemShut
  {NoStop}%
\bibitem [{\citenamefont {Peyronel}\ \emph {et~al.}(2012)\citenamefont
  {Peyronel}, \citenamefont {Firstenberg}, \citenamefont {Liang}, \citenamefont
  {Hofferberth}, \citenamefont {Gorshkov}, \citenamefont {Pohl}, \citenamefont
  {Lukin},\ and\ \citenamefont {Vuleti{\'c}}}]{peyronel2012quantum}%
  \BibitemOpen
  \bibfield  {author} {\bibinfo {author} {\bibfnamefont {T.}~\bibnamefont
  {Peyronel}}, \bibinfo {author} {\bibfnamefont {O.}~\bibnamefont
  {Firstenberg}}, \bibinfo {author} {\bibfnamefont {Q.-Y.}\ \bibnamefont
  {Liang}}, \bibinfo {author} {\bibfnamefont {S.}~\bibnamefont {Hofferberth}},
  \bibinfo {author} {\bibfnamefont {A.~V.}\ \bibnamefont {Gorshkov}}, \bibinfo
  {author} {\bibfnamefont {T.}~\bibnamefont {Pohl}}, \bibinfo {author}
  {\bibfnamefont {M.~D.}\ \bibnamefont {Lukin}},\ and\ \bibinfo {author}
  {\bibfnamefont {V.}~\bibnamefont {Vuleti{\'c}}},\ }\bibfield  {title}
  {\bibinfo {title} {Quantum nonlinear optics with single photons enabled by
  strongly interacting atoms},\ }\href@noop {} {\bibfield  {journal} {\bibinfo
  {journal} {Nature}\ }\textbf {\bibinfo {volume} {488}},\ \bibinfo {pages}
  {57} (\bibinfo {year} {2012})}\BibitemShut {NoStop}%
\bibitem [{\citenamefont {Busche}\ \emph {et~al.}(2017)\citenamefont {Busche},
  \citenamefont {Huillery}, \citenamefont {Ball}, \citenamefont {Ilieva},
  \citenamefont {Jones},\ and\ \citenamefont {Adams}}]{busche2017contactless}%
  \BibitemOpen
  \bibfield  {author} {\bibinfo {author} {\bibfnamefont {H.}~\bibnamefont
  {Busche}}, \bibinfo {author} {\bibfnamefont {P.}~\bibnamefont {Huillery}},
  \bibinfo {author} {\bibfnamefont {S.~W.}\ \bibnamefont {Ball}}, \bibinfo
  {author} {\bibfnamefont {T.}~\bibnamefont {Ilieva}}, \bibinfo {author}
  {\bibfnamefont {M.~P.~A.}\ \bibnamefont {Jones}},\ and\ \bibinfo {author}
  {\bibfnamefont {C.~S.}\ \bibnamefont {Adams}},\ }\bibfield  {title} {\bibinfo
  {title} {Contactless nonlinear optics mediated by long-range rydberg
  interactions},\ }\href@noop {} {\bibfield  {journal} {\bibinfo  {journal}
  {Nat. Phys.}\ }\textbf {\bibinfo {volume} {13}},\ \bibinfo {pages} {655}
  (\bibinfo {year} {2017})}\BibitemShut {NoStop}%
\bibitem [{\citenamefont {Tiarks}\ \emph {et~al.}(2019)\citenamefont {Tiarks},
  \citenamefont {Schmidt-Eberle}, \citenamefont {Stolz}, \citenamefont
  {Rempe},\ and\ \citenamefont {D{\"u}rr}}]{tiarks2019photongate}%
  \BibitemOpen
  \bibfield  {author} {\bibinfo {author} {\bibfnamefont {D.}~\bibnamefont
  {Tiarks}}, \bibinfo {author} {\bibfnamefont {S.}~\bibnamefont
  {Schmidt-Eberle}}, \bibinfo {author} {\bibfnamefont {T.}~\bibnamefont
  {Stolz}}, \bibinfo {author} {\bibfnamefont {G.}~\bibnamefont {Rempe}},\ and\
  \bibinfo {author} {\bibfnamefont {S.}~\bibnamefont {D{\"u}rr}},\ }\bibfield
  {title} {\bibinfo {title} {A photon–photon quantum gate based on rydberg
  interactions},\ }\href@noop {} {\bibfield  {journal} {\bibinfo  {journal}
  {Nat. Phys.}\ }\textbf {\bibinfo {volume} {15}},\ \bibinfo {pages} {124}
  (\bibinfo {year} {2019})}\BibitemShut {NoStop}%
\bibitem [{\citenamefont {Mohapatra}\ \emph {et~al.}(2008)\citenamefont
  {Mohapatra}, \citenamefont {Bason}, \citenamefont {Butscher}, \citenamefont
  {Weatherill},\ and\ \citenamefont {Adams}}]{mohapatra2008giant}%
  \BibitemOpen
  \bibfield  {author} {\bibinfo {author} {\bibfnamefont {A.~K.}\ \bibnamefont
  {Mohapatra}}, \bibinfo {author} {\bibfnamefont {M.~G.}\ \bibnamefont
  {Bason}}, \bibinfo {author} {\bibfnamefont {B.}~\bibnamefont {Butscher}},
  \bibinfo {author} {\bibfnamefont {K.~J.}\ \bibnamefont {Weatherill}},\ and\
  \bibinfo {author} {\bibfnamefont {C.~S.}\ \bibnamefont {Adams}},\ }\bibfield
  {title} {\bibinfo {title} {A giant electro-optic effect using polarizable
  dark states},\ }\href@noop {} {\bibfield  {journal} {\bibinfo  {journal}
  {np}\ }\textbf {\bibinfo {volume} {4}},\ \bibinfo {pages} {890} (\bibinfo
  {year} {2008})}\BibitemShut {NoStop}%
\bibitem [{\citenamefont {Sedlacek}\ \emph {et~al.}(2012)\citenamefont
  {Sedlacek}, \citenamefont {Schwettmann}, \citenamefont {K{\"u}bler},
  \citenamefont {L{\"o}w}, \citenamefont {Pfau},\ and\ \citenamefont
  {Shaffer}}]{sedlacek2012microwave}%
  \BibitemOpen
  \bibfield  {author} {\bibinfo {author} {\bibfnamefont {J.~A.}\ \bibnamefont
  {Sedlacek}}, \bibinfo {author} {\bibfnamefont {A.}~\bibnamefont
  {Schwettmann}}, \bibinfo {author} {\bibfnamefont {H.}~\bibnamefont
  {K{\"u}bler}}, \bibinfo {author} {\bibfnamefont {R.}~\bibnamefont {L{\"o}w}},
  \bibinfo {author} {\bibfnamefont {T.}~\bibnamefont {Pfau}},\ and\ \bibinfo
  {author} {\bibfnamefont {J.~P.}\ \bibnamefont {Shaffer}},\ }\bibfield
  {title} {\bibinfo {title} {Microwave electrometry with rydberg atoms in a
  vapour cell using bright atomic resonances},\ }\href@noop {} {\bibfield
  {journal} {\bibinfo  {journal} {Nat. Phys.}\ }\textbf {\bibinfo {volume}
  {8}},\ \bibinfo {pages} {819} (\bibinfo {year} {2012})}\BibitemShut {NoStop}%
\bibitem [{\citenamefont {Wade}\ \emph {et~al.}(2017)\citenamefont {Wade},
  \citenamefont {\u{S}ibali\'{c}}, \citenamefont {de~Melo}, \citenamefont
  {Kondo}, \citenamefont {Adams},\ and\ \citenamefont {Weatherill}}]{Wade2017}%
  \BibitemOpen
  \bibfield  {author} {\bibinfo {author} {\bibfnamefont {C.~G.}\ \bibnamefont
  {Wade}}, \bibinfo {author} {\bibfnamefont {N.}~\bibnamefont
  {\u{S}ibali\'{c}}}, \bibinfo {author} {\bibfnamefont {N.~R.}\ \bibnamefont
  {de~Melo}}, \bibinfo {author} {\bibfnamefont {J.~M.}\ \bibnamefont {Kondo}},
  \bibinfo {author} {\bibfnamefont {C.~S.}\ \bibnamefont {Adams}},\ and\
  \bibinfo {author} {\bibfnamefont {K.~J.}\ \bibnamefont {Weatherill}},\
  }\bibfield  {title} {\bibinfo {title} {Real-time near-field terahertz imaging
  with atomic optical fluorescence},\ }\href
  {http://dx.doi.org/10.1038/nphoton.2016.214} {\bibfield  {journal} {\bibinfo
  {journal} {Nat. Photon.}\ }\textbf {\bibinfo {volume} {11}},\ \bibinfo
  {pages} {40} (\bibinfo {year} {2017})}\BibitemShut {NoStop}%
\bibitem [{\citenamefont {Holloway}\ \emph {et~al.}(2017)\citenamefont
  {Holloway}, \citenamefont {Simons}, \citenamefont {Gordon}, \citenamefont
  {Wilson}, \citenamefont {Cooke}, \citenamefont {Anderson},\ and\
  \citenamefont {Raithel}}]{holloway2017rfmetrology}%
  \BibitemOpen
  \bibfield  {author} {\bibinfo {author} {\bibfnamefont {C.~L.}\ \bibnamefont
  {Holloway}}, \bibinfo {author} {\bibfnamefont {M.~T.}\ \bibnamefont
  {Simons}}, \bibinfo {author} {\bibfnamefont {J.~A.}\ \bibnamefont {Gordon}},
  \bibinfo {author} {\bibfnamefont {P.~F.}\ \bibnamefont {Wilson}}, \bibinfo
  {author} {\bibfnamefont {C.~M.}\ \bibnamefont {Cooke}}, \bibinfo {author}
  {\bibfnamefont {D.~A.}\ \bibnamefont {Anderson}},\ and\ \bibinfo {author}
  {\bibfnamefont {G.}~\bibnamefont {Raithel}},\ }\bibfield  {title} {\bibinfo
  {title} {Atom-based rf electric field metrology: From self-calibrated
  measurements to subwavelength and near-field imaging},\ }\href@noop {}
  {\bibfield  {journal} {\bibinfo  {journal} {IEEE Transactions on
  Electromagnetic Compatibility}\ }\textbf {\bibinfo {volume} {59}},\ \bibinfo
  {pages} {717} (\bibinfo {year} {2017})}\BibitemShut {NoStop}%
\bibitem [{\citenamefont {Deb}\ and\ \citenamefont
  {Kj{\ae}rgaard}(2018)}]{deb2018radio}%
  \BibitemOpen
  \bibfield  {author} {\bibinfo {author} {\bibfnamefont {A.}~\bibnamefont
  {Deb}}\ and\ \bibinfo {author} {\bibfnamefont {N.}~\bibnamefont
  {Kj{\ae}rgaard}},\ }\bibfield  {title} {\bibinfo {title} {Radio-over-fiber
  using an optical antenna based on rydberg states of atoms},\ }\href@noop {}
  {\bibfield  {journal} {\bibinfo  {journal} {Applied Physics Letters}\
  }\textbf {\bibinfo {volume} {112}},\ \bibinfo {pages} {211106} (\bibinfo
  {year} {2018})}\BibitemShut {NoStop}%
\bibitem [{\citenamefont {Meyer}\ \emph {et~al.}(2018)\citenamefont {Meyer},
  \citenamefont {Cox}, \citenamefont {Fatemi},\ and\ \citenamefont
  {Kunz}}]{meyer2018digital}%
  \BibitemOpen
  \bibfield  {author} {\bibinfo {author} {\bibfnamefont {D.~H.}\ \bibnamefont
  {Meyer}}, \bibinfo {author} {\bibfnamefont {K.~C.}\ \bibnamefont {Cox}},
  \bibinfo {author} {\bibfnamefont {F.~K.}\ \bibnamefont {Fatemi}},\ and\
  \bibinfo {author} {\bibfnamefont {P.~D.}\ \bibnamefont {Kunz}},\ }\bibfield
  {title} {\bibinfo {title} {Digital communication with rydberg atoms and
  amplitude-modulated microwave fields},\ }\href@noop {} {\bibfield  {journal}
  {\bibinfo  {journal} {Applied Physics Letters}\ }\textbf {\bibinfo {volume}
  {112}},\ \bibinfo {pages} {211108} (\bibinfo {year} {2018})}\BibitemShut
  {NoStop}%
\bibitem [{\citenamefont {Jing}\ \emph {et~al.}(2020)\citenamefont {Jing},
  \citenamefont {Hu}, \citenamefont {Ma}, \citenamefont {Zhang}, \citenamefont
  {Zhang}, \citenamefont {Xiao},\ and\ \citenamefont {Jia}}]{jing2020}%
  \BibitemOpen
  \bibfield  {author} {\bibinfo {author} {\bibfnamefont {M.}~\bibnamefont
  {Jing}}, \bibinfo {author} {\bibfnamefont {Y.}~\bibnamefont {Hu}}, \bibinfo
  {author} {\bibfnamefont {J.}~\bibnamefont {Ma}}, \bibinfo {author}
  {\bibfnamefont {H.}~\bibnamefont {Zhang}}, \bibinfo {author} {\bibfnamefont
  {L.}~\bibnamefont {Zhang}}, \bibinfo {author} {\bibfnamefont
  {L.}~\bibnamefont {Xiao}},\ and\ \bibinfo {author} {\bibfnamefont
  {S.}~\bibnamefont {Jia}},\ }\bibfield  {title} {\bibinfo {title} {Atomic
  superheterodyne receiver based on microwave-dressed rydberg spectroscopy},\
  }\href@noop {} {\bibfield  {journal} {\bibinfo  {journal} {Nat. Phys.}\
  }\textbf {\bibinfo {volume} {16}},\ \bibinfo {pages} {911} (\bibinfo {year}
  {2020})}\BibitemShut {NoStop}%
\bibitem [{\citenamefont {Downes}\ \emph {et~al.}(2020)\citenamefont {Downes},
  \citenamefont {MacKellar}, \citenamefont {Whiting}, \citenamefont
  {Bourgenot}, \citenamefont {Adams}, \citenamefont {1},\ and\ \citenamefont
  {Weatherill}}]{downes2020}%
  \BibitemOpen
  \bibfield  {author} {\bibinfo {author} {\bibfnamefont {L.~A.}\ \bibnamefont
  {Downes}}, \bibinfo {author} {\bibfnamefont {A.~R.}\ \bibnamefont
  {MacKellar}}, \bibinfo {author} {\bibfnamefont {D.~J.}\ \bibnamefont
  {Whiting}}, \bibinfo {author} {\bibfnamefont {C.}~\bibnamefont {Bourgenot}},
  \bibinfo {author} {\bibfnamefont {C.~S.}\ \bibnamefont {Adams}}, \bibinfo
  {author} {\bibnamefont {1}},\ and\ \bibinfo {author} {\bibfnamefont {K.~J.}\
  \bibnamefont {Weatherill}},\ }\bibfield  {title} {\bibinfo {title}
  {Full-field terahertz imaging at kilohertz frame rates using atomic vapor},\
  }\href@noop {} {\bibfield  {journal} {\bibinfo  {journal} {Phys. Rev. X}\
  }\textbf {\bibinfo {volume} {10}},\ \bibinfo {pages} {011027} (\bibinfo
  {year} {2020})}\BibitemShut {NoStop}%
\bibitem [{\citenamefont {Lam}\ \emph {et~al.}(2021)\citenamefont {Lam},
  \citenamefont {Pal}, \citenamefont {Vogt}, \citenamefont {Kiffner},\ and\
  \citenamefont {Li}}]{lam2021directional}%
  \BibitemOpen
  \bibfield  {author} {\bibinfo {author} {\bibfnamefont {M.}~\bibnamefont
  {Lam}}, \bibinfo {author} {\bibfnamefont {S.~B.}\ \bibnamefont {Pal}},
  \bibinfo {author} {\bibfnamefont {T.}~\bibnamefont {Vogt}}, \bibinfo {author}
  {\bibfnamefont {M.}~\bibnamefont {Kiffner}},\ and\ \bibinfo {author}
  {\bibfnamefont {W.}~\bibnamefont {Li}},\ }\bibfield  {title} {\bibinfo
  {title} {Directional thz generation in hot rb vapor excited to a rydberg
  state},\ }\href@noop {} {\bibfield  {journal} {\bibinfo  {journal} {Optics
  Letters}\ }\textbf {\bibinfo {volume} {46}},\ \bibinfo {pages} {1017}
  (\bibinfo {year} {2021})}\BibitemShut {NoStop}%
\bibitem [{\citenamefont {Han}\ \emph {et~al.}(2018)\citenamefont {Han},
  \citenamefont {Vogt}, \citenamefont {Gross}, \citenamefont {Jaksch},
  \citenamefont {Kiffner},\ and\ \citenamefont {Li}}]{han2018coherent}%
  \BibitemOpen
  \bibfield  {author} {\bibinfo {author} {\bibfnamefont {J.}~\bibnamefont
  {Han}}, \bibinfo {author} {\bibfnamefont {T.}~\bibnamefont {Vogt}}, \bibinfo
  {author} {\bibfnamefont {C.}~\bibnamefont {Gross}}, \bibinfo {author}
  {\bibfnamefont {D.}~\bibnamefont {Jaksch}}, \bibinfo {author} {\bibfnamefont
  {M.}~\bibnamefont {Kiffner}},\ and\ \bibinfo {author} {\bibfnamefont
  {W.}~\bibnamefont {Li}},\ }\bibfield  {title} {\bibinfo {title} {Coherent
  microwave-to-optical conversion via six-wave mixing in rydberg atoms},\
  }\href@noop {} {\bibfield  {journal} {\bibinfo  {journal} {Physical review
  letters}\ }\textbf {\bibinfo {volume} {120}},\ \bibinfo {pages} {093201}
  (\bibinfo {year} {2018})}\BibitemShut {NoStop}%
\bibitem [{\citenamefont {Tu}\ \emph {et~al.}(2022)\citenamefont {Tu},
  \citenamefont {Liao}, \citenamefont {Zhang}, \citenamefont {Liu},
  \citenamefont {Zheng}, \citenamefont {Yang}, \citenamefont {Zhang},
  \citenamefont {Yan},\ and\ \citenamefont {Zhu}}]{tu2022}%
  \BibitemOpen
  \bibfield  {author} {\bibinfo {author} {\bibfnamefont {H.-T.}\ \bibnamefont
  {Tu}}, \bibinfo {author} {\bibfnamefont {K.-Y.}\ \bibnamefont {Liao}},
  \bibinfo {author} {\bibfnamefont {Z.-X.}\ \bibnamefont {Zhang}}, \bibinfo
  {author} {\bibfnamefont {X.-H.}\ \bibnamefont {Liu}}, \bibinfo {author}
  {\bibfnamefont {S.-Y.}\ \bibnamefont {Zheng}}, \bibinfo {author}
  {\bibfnamefont {S.-Z.}\ \bibnamefont {Yang}}, \bibinfo {author}
  {\bibfnamefont {X.-D.}\ \bibnamefont {Zhang}}, \bibinfo {author}
  {\bibfnamefont {H.}~\bibnamefont {Yan}},\ and\ \bibinfo {author}
  {\bibfnamefont {S.-L.}\ \bibnamefont {Zhu}},\ }\bibfield  {title} {\bibinfo
  {title} {High-efficiency coherent microwave-to-optics conversion via
  off-resonant scattering},\ }\href@noop {} {\bibfield  {journal} {\bibinfo
  {journal} {Nat. Photon.}\ }\textbf {\bibinfo {volume} {16}},\ \bibinfo
  {pages} {291} (\bibinfo {year} {2022})}\BibitemShut {NoStop}%
\bibitem [{\citenamefont {Carr}\ \emph {et~al.}(2012)\citenamefont {Carr},
  \citenamefont {Tanasittikosol}, \citenamefont {Sargsyan}, \citenamefont
  {Sarkisyan}, \citenamefont {Adams},\ and\ \citenamefont
  {Weatherill}}]{carr2012three}%
  \BibitemOpen
  \bibfield  {author} {\bibinfo {author} {\bibfnamefont {C.}~\bibnamefont
  {Carr}}, \bibinfo {author} {\bibfnamefont {M.}~\bibnamefont
  {Tanasittikosol}}, \bibinfo {author} {\bibfnamefont {A.}~\bibnamefont
  {Sargsyan}}, \bibinfo {author} {\bibfnamefont {D.}~\bibnamefont {Sarkisyan}},
  \bibinfo {author} {\bibfnamefont {C.~S.}\ \bibnamefont {Adams}},\ and\
  \bibinfo {author} {\bibfnamefont {K.~J.}\ \bibnamefont {Weatherill}},\
  }\bibfield  {title} {\bibinfo {title} {Three-photon electromagnetically
  induced transparency using rydberg states},\ }\href@noop {} {\bibfield
  {journal} {\bibinfo  {journal} {Opt. Lett.}\ }\textbf {\bibinfo {volume}
  {37}},\ \bibinfo {pages} {3858} (\bibinfo {year} {2012})}\BibitemShut
  {NoStop}%
\bibitem [{\citenamefont {Thaicharoen}\ \emph {et~al.}(2019)\citenamefont
  {Thaicharoen}, \citenamefont {Moore}, \citenamefont {Anderson}, \citenamefont
  {Powel}, \citenamefont {Peterson},\ and\ \citenamefont
  {Raithel}}]{thaicharoen2019}%
  \BibitemOpen
  \bibfield  {author} {\bibinfo {author} {\bibfnamefont {N.}~\bibnamefont
  {Thaicharoen}}, \bibinfo {author} {\bibfnamefont {K.~R.}\ \bibnamefont
  {Moore}}, \bibinfo {author} {\bibfnamefont {D.~A.}\ \bibnamefont {Anderson}},
  \bibinfo {author} {\bibfnamefont {R.~C.}\ \bibnamefont {Powel}}, \bibinfo
  {author} {\bibfnamefont {E.}~\bibnamefont {Peterson}},\ and\ \bibinfo
  {author} {\bibfnamefont {G.}~\bibnamefont {Raithel}},\ }\bibfield  {title}
  {\bibinfo {title} {Electromagnetically induced transparency, absorption, and
  microwave-field sensing in a rb vapor cell with a three-color all-infrared
  laser system},\ }\href@noop {} {\bibfield  {journal} {\bibinfo  {journal}
  {Phys. Rev. A}\ }\textbf {\bibinfo {volume} {100}},\ \bibinfo {pages}
  {063427} (\bibinfo {year} {2019})}\BibitemShut {NoStop}%
\bibitem [{\citenamefont {Ripka}\ \emph {et~al.}(2021)\citenamefont {Ripka},
  \citenamefont {Amarloo}, \citenamefont {Erskine}, \citenamefont {Liu},
  \citenamefont {Ramirez-Serrano}, \citenamefont {Keaveny}, \citenamefont
  {Gillet}, \citenamefont {Kubler},\ and\ \citenamefont {Shaffer}}]{ripka2021}%
  \BibitemOpen
  \bibfield  {author} {\bibinfo {author} {\bibfnamefont {F.}~\bibnamefont
  {Ripka}}, \bibinfo {author} {\bibfnamefont {H.}~\bibnamefont {Amarloo}},
  \bibinfo {author} {\bibfnamefont {J.}~\bibnamefont {Erskine}}, \bibinfo
  {author} {\bibfnamefont {C.}~\bibnamefont {Liu}}, \bibinfo {author}
  {\bibfnamefont {J.}~\bibnamefont {Ramirez-Serrano}}, \bibinfo {author}
  {\bibfnamefont {J.}~\bibnamefont {Keaveny}}, \bibinfo {author} {\bibfnamefont
  {G.}~\bibnamefont {Gillet}}, \bibinfo {author} {\bibfnamefont
  {H.}~\bibnamefont {Kubler}},\ and\ \bibinfo {author} {\bibfnamefont {J.~P.}\
  \bibnamefont {Shaffer}},\ }\bibfield  {title} {\bibinfo {title}
  {Application-driven problems in rydberg atom electrometry},\ }\href@noop {}
  {\bibfield  {journal} {\bibinfo  {journal} {Proc. SPIE}\ }\textbf {\bibinfo
  {volume} {11700}},\ \bibinfo {pages} {117002} (\bibinfo {year}
  {2021})}\BibitemShut {NoStop}%
\bibitem [{\citenamefont {Akulshin}\ \emph {et~al.}(2009)\citenamefont
  {Akulshin}, \citenamefont {McLean}, \citenamefont {Sidorov},\ and\
  \citenamefont {Hannaford}}]{Akulshin:09}%
  \BibitemOpen
  \bibfield  {author} {\bibinfo {author} {\bibfnamefont {A.~M.}\ \bibnamefont
  {Akulshin}}, \bibinfo {author} {\bibfnamefont {R.~J.}\ \bibnamefont
  {McLean}}, \bibinfo {author} {\bibfnamefont {A.~I.}\ \bibnamefont
  {Sidorov}},\ and\ \bibinfo {author} {\bibfnamefont {P.}~\bibnamefont
  {Hannaford}},\ }\bibfield  {title} {\bibinfo {title} {Coherent and collimated
  blue light generated by four-wave mixing in rb vapour},\ }\href
  {https://doi.org/10.1364/OE.17.022861} {\bibfield  {journal} {\bibinfo
  {journal} {Opt. Express}\ }\textbf {\bibinfo {volume} {17}},\ \bibinfo
  {pages} {22861} (\bibinfo {year} {2009})}\BibitemShut {NoStop}%
\bibitem [{\citenamefont {Vernier}\ \emph {et~al.}(2010)\citenamefont
  {Vernier}, \citenamefont {Franke-Arnold}, \citenamefont {Riis},\ and\
  \citenamefont {Arnold}}]{Vernier:10}%
  \BibitemOpen
  \bibfield  {author} {\bibinfo {author} {\bibfnamefont {A.}~\bibnamefont
  {Vernier}}, \bibinfo {author} {\bibfnamefont {S.}~\bibnamefont
  {Franke-Arnold}}, \bibinfo {author} {\bibfnamefont {E.}~\bibnamefont
  {Riis}},\ and\ \bibinfo {author} {\bibfnamefont {A.~S.}\ \bibnamefont
  {Arnold}},\ }\bibfield  {title} {\bibinfo {title} {Enhanced frequency
  up-conversion in rb vapor},\ }\href {https://doi.org/10.1364/OE.18.017020}
  {\bibfield  {journal} {\bibinfo  {journal} {Opt. Express}\ }\textbf {\bibinfo
  {volume} {18}},\ \bibinfo {pages} {17020} (\bibinfo {year}
  {2010})}\BibitemShut {NoStop}%
\end{thebibliography}
\end{document}